\documentstyle[pra,aps,epsfig,twocolumn]{revtex}

\def\be{\begin{equation}}
\def\ee{\end{equation}}
\def\bea{\begin{eqnarray}}
\def\eea{\end{eqnarray}}
\def\bma{\begin{mathletters}}
\def\ema{\end{mathletters}}
\def\C{\hbox{$\mit /$\kern-.6em$\mit C$}}
\def\one{\hbox{$\mit I$\kern-.6em$\mit I$}}

\begin{document}
\title{Generation and interaction of solitons in Bose-Einstein
Condensates}
\author{
S. Burger$^{1}
\footnote{Current address: L.E.N.S., 50125 Firenze, Italia}$, 
L. D. Carr$^{2}$, 
P. \"Ohberg$^{3}$,  
K. Sengstock$^{1}$, and A. Sanpera$^{4}$}
\address{$^{1}$ Institut f\"ur Quantenoptik, Universit\"at  Hannover, Germany\\}
\address{$^{2}$Department of Physics, University of Washington, 
         Seattle, WA 98195-1560, USA\\}
\address{$^{3 }$ School of Physics and Astronomy, University of St Andrews, North Haugh, 
St Andrews, Fife KY16 9SS, Scotland\\}
\address{$^{4}$Institut f\"ur Theoretische Physik, Universit\"at Hannover, 30167 Hannover, Germany\\}

\maketitle
\date{\today}

\begin{abstract}

Generation, interaction and detection of dark solitons in Bose-Einstein
condensates is considered. In particular, we focus on the dynamics
resulting from  phase imprinting and density engineering. 
The generation of soliton pairs  as well as their interaction is also
considered.    Finally, motivated by the recent experimental results of
Cornish {\it {et al.}}  (Phys. Rev Lett. {\bf 85}, 1795, 2000), 
we analyze the stability of dark solitons under changes 
of the scattering  length and thereby demonstrate a new way to detect
them.  Our theoretical and numerical results compare well 
with the existing experimental ones and provide guidance for future
experiments.
\end{abstract}


\section{Introduction}

Bose-Einstein condensates (BEC's) offer a unique possibility 
of studying   nonlinear effects using matter waves. 
This has been spectacularly shown in the recent  BEC experiments 
which demonstrate, among other things, the possibility of four wave mixing
\cite{deng99}, the creation of topological structures such as
vortices ~\cite{JILA,ENS}, the creation of
solitons~\cite{bur99,dens00,anderson00}, as well as other
demonstrations of the superfluid character~\cite{arlt}.

Solitons are one dimensional waves that propagate without spreading 
in a nonlinear medium.
Their shape remains unaltered after interacting with other solitons.
The nonlinear Schr\"odinger equation, which accurately describes
dilute BEC's at zero temperature, supports soliton solutions for
attractive as well as for repulsive 2--body
interactions\cite{zakharov72,zakharov73}.  These solutions 
correspond to macroscopically excited states of the mean 
field of the condensate. For a single component condensate, the 3D 
time-dependent nonlinear
Schr\"odinger equation, also known as the Gross-Pitaevskii
equation (GPE), reads 
\be
i\hbar \frac{\partial}{\partial t}\Psi(\vec{r},t)=
\left \{ -\frac{\hbar^2}{2m}{{\nabla}^2}+ V(\vec{r})+
g|\Psi(\vec{r},t)|^2 \right \}\Psi(\vec{r},t).
\label{3DGP}
\ee
Here $g=4\pi\hbar^2 a/m$, where $a$ corresponds to the 
$s$-wave scattering
length for binary collisions between atoms,
$m$ refers to the mass of the atoms and $V$ to the trap potential.
For repulsive interactions ($a>0$), 
solitons are characterized by a local density minimum
together with a sharp phase gradient of the wave function at the
position of the minimum. 
In this case, 
the nonlinear effective mean field potential term $g|\Psi(\vec{r},t)|^2$ 
balances the dispersion of the wavefunction caused by the kinetic energy.
Because there is a notch in the density these solutions are
termed {\it dark} solitons. 
Only when the density of the condensate exactly vanishes at the density
minimum, is the resulting soliton   stationary.  Such
a standing soliton has exactly a  phase jump of $\pi$ between the two parts
of the condensate connected by it and its velocity is zero. 
In general, the local minimum density can range from maximal  to zero
depth  with the associated soliton velocity, $\dot{q}$, ranging from zero to  the
speed of sound, $c_s$, in the condensate, i.e. 
$0\leq \dot{q}\leq c_{s}=\sqrt{n_0 g/m}$, where $n_0$ refers 
to the mean density in the condensate. 
In a homogeneous 1D condensate, a soliton solution  
can  be analytically obtained in an elegant way by using the 
inverse scattering method\cite{zakharov73}. 
The wavefunction corresponding
to a  dark soliton located at $q$ 
propagating along the $z$-axis with speed  $v$ 
is described by   
\begin{eqnarray}
&&\Psi_{dark}(z,t)=\sqrt{n_0}\left \{ i\frac
{v}{c_s}+\right. \nonumber \\ &&\left. \sqrt{1-\frac{v^2}{c_s^2}}
\tanh \left [\sqrt{1-\frac{v^2}{c_s^2}} \,\frac{(z-v\,t)}{\sqrt{2}\,l_0}\right ]\right\}\,e^{-ign_0t/\hbar}\,.
\label{darksol}
\end{eqnarray}
In the case of a dark soliton  
propagating along the  $z$-axis in an elongated 3D condensate, 
$z$ refers to the position of the notch ($x-y$) plane,  
and $v$ the constant velocity of such a plane   with respect
to a stationary background.
The soliton size is of the order of twice the healing 
length $l_0=\hbar/\sqrt{m g n_0}$.

In contrast, for condensates with attractive interactions  ($a<0$) 
soliton solutions are characterized by a maximum 
in the density profile without any phase jump across it. 
These solitons are termed {\it bright} solitons, and the solution
corresponding to a bright soliton reads:   
\be 
\Psi_{bright}=\sqrt{n_0}\,{\rm sech} \left (\frac{z-v\,t}{l_0}\right )
e^{-i2mvz/\hbar}e^{-ign_0 t/\hbar}.
\label{bright}
\ee
The ground state of a condensate with
attractive interactions and sufficiently large nonlinearity, i.e., $g >
g_{min}$, is, in fact, a bright soliton \cite{linconatrac}.

Dark and bright solitons described by the nonlinear Schr\"odinger 
equation have been extensively studied in the context of nonlinear optics
(see ~\cite{Kiv98} and references therein). 
In condensates only dark solitons have very recently 
been observed as nonlinear matter waves \cite{bur99,dens00,anderson00}.
A crucial difference  between optical and matter wave solitons
appears at first glance. While optical solitons are created in 
optical guides, i.e. in a cylindrical medium which is 
a priori unbounded, a condensate is always confined in a trap.
To what extent the boundary conditions affect the properties 
and stability of the solitons has been the subject of recent theoretical
studies (see e.g. \cite{ang00}).

The literature concerning 
bright solitons in matter waves is not very extensive
\cite{zakharov72,lenz94,meystre,linconatrac,busch01}.
Condensates with attractive interactions in 2 and 3D 
are unstable objects that collapse very rapidly when the particle number 
becomes too large\cite{bradley97,ruprecht95}. 
Therefore a bright soliton in a condensate with attractive interactions
is  also an unstable object.  However, since quasi-1D condensates
with attractive interactions are stable, bright solitons
could be generated in them
\cite{carr221}.  Alternatively, an interesting approach 
based on vector solitons in
two-component condensates (with repulsive interactions) can also 
be used to to study bright solitons in 2 and 3D condensates.
By creating  a dark soliton in one of the species  
one can induce a bright soliton-like structure in the other 
one~\cite{anderson00,busch01,patrik01} which fills the minimum of the
first species.  Since a condensate with positive scattering length is
stable,  the bright soliton thus created is no longer limited by the
collapse  of the condensate. Finally, 
it has been recently shown that 
with the well-controlled use of Feshbach resonances in 
$^{85}$Rb, it is possible to make a condensate with attractive interactions
in a very controllable way\cite{cornish00,scat00}. 
This opens a new way to study bright solitons in matter waves.
 
The study of matter wave solitons, both experimentally and theoretically, 
principally involves three different aspects: their generation, 
coherent evolution including coherent effects during 
detection,  and incoherent evolution and dissipation. 
So far the observed solitons in one component BEC's  
have been generated by the method of phase imprinting. 
This method, originally proposed to
generate vortices, has become a very efficient tool to engineer the phase 
in condensates~\cite{dobrek99,dobrek01}. 
Optimization of the phase imprinting method has been recently discussed 
by Carr {\it{et al.}} \cite{carr00}, where initially not only the phase 
but also the density is properly engineered. 
A proper combination of both effects in a quasi-homogeneous condensate 
produces a stable, standing soliton whose properties can be used to test
 fundamental aspects of many-body theory such as quantum and thermal
fluctuations.

The coherent evolution of solitons refers to the evolution before
dissipation takes place. While a condensate
has a lifetime on the order of seconds, the recent experimental results 
have shown that for solitons the lifetime is on the order of
15 milliseconds (in one component condensates). 
For shorter times the  Gross-Pitaevskii equation provides an accurate 
description of the soliton dynamics~\cite{bur99,dens00,anderson00}.
In homogeneous systems and in the absence of dissipation and/or thermal 
phonons, solitons maintain a constant velocity. 
In elongated harmonic traps a dark soliton 
(and in 3D the nodal plane of the soliton) 
oscillates in the trap with a frequency $\Omega=\omega_a/ \sqrt{2}$ 
where $\omega_a$ corresponds to the axial trap frequency \cite{ang00}.  
Coherent evolution also concerns the generation and interaction of
pairs  of solitons. This coherent soliton evolution has been 
recently studied in the context of two-component condensates, 
where new and rich dynamics appears \cite{patrik01}. 

Dissipative effects include both dynamical  
and thermal instabilities. Dynamical instabilities are 
due to the fact that solitons are indeed one dimensional 
objects. When embedded in higher dimensions 
(i.e. 2 or 3D condensates) their stability strongly depends 
on the geometry of the condensate. For sufficiently elongated traps 
with a high transverse confinement it is  not possible to excite 
the transverse  modes  of the trap and the soliton is then dynamically 
stable. On the contrary, for a looser transverse confinement 
the transverse modes can be excited, making 
the soliton plane bend and undergo a snake 
instability. As a result, the soliton 
decays into phonons and  more stable structures such as vortices and vortex rings\cite{anderson00,donnelly,martikainen},   or even more exotic objects such as svortices\cite{brand}. 
Depending on the trap geometry, dynamical instabilities 
may occur on a shorter time scale than thermal instabilities. 
One should point out here that the dissipatory behavior of solitons 
corresponding to dynamical instabilities can also be described, using 
the Gross-Pitaevskii equation, as coherent but unstable evolution
\cite{fed99,mur99}. 

Thermal instabilities appear due to the fact
that solitons are collective {\it excited} states of the mean field of
the condensate, and, therefore, decay into the ground state 
within a finite time.
The dissipation consists in scattering of phonons on the soliton's 
notch plane. Studies of dissipation, which are related to the
interaction of a soliton with a thermal cloud, 
demand   one to go beyond the GPE and use the Bogoliubov-deGennes
  equations
which describe collective modes such as phonons. 
Since a soliton can be regarded as a particle with negative mass, 
such scattering accelerates the soliton until it reaches the effective sound 
velocity and vanishes. This scenario has been described
by  Fedichev {\it{et al.}}~\cite{fed99}, and has been recently studied for 
3D solitons in elongated traps by Muryshev {\it {et al.}}\cite{mur00}.
Both theory and experiment indicate that the life time of the soliton
due to  thermodynamic instability is of the order of 15 ms. 
This contrasts with the lifetime of vortices which is of the order of a few
 seconds \cite{JILA,ENS}.

This paper focuses on the generation and coherent evolution 
of dark solitons in   one-component condensates in elongated traps.  
It is  organized as follows: 
Section II addresses the issue of the generation of solitons. 
In Section III the problem of soliton detection is investigated 
and the dynamics concerning the opening of the trap are studied. 
Section IV is devoted to the interaction between solitons. 
We discuss therein the experimental conditions under which the effects 
due to interactions can be
observed. 
In the latter, we study the stability of a soliton when the scattering
length of the condensate is changed in a controllable way. This study is
stimulated by the recent experiments in $^{85}$Rb by Cornish {\it{et al.}}
\cite{cornish00,scat00}.
Finally we present our conclusions in Section V.

\section{Generation of dark solitons}

In this section we focus on the generation of dark solitons in one
component Bose-Einstein condensates by three distinct
methods: (a) phase imprinting, (b) 
density engineering, i.e., non-adiabatic changes of the potential
confining the condensate and (c) combination of the previous two methods. 

As already mentioned in the introduction, even for thermodynamically
unstable states such as solitons the GPE provides an excellent tool 
with which to describe the coherent dynamics on the relevant timescale.
To avoid  the effect of dynamical instabilities we restrict our analysis 
to the following configurations: 
(a)   cigar-shaped condensates with a high aspect ratio and 
(b) quasi-1D condensates. These quasi-1D condensates can be obtained
for both harmonic and box-like   potentials. For the former, the radial
confinement frequency is required to be much larger than the mean field
interaction between particles\cite{mur00}. 
For the latter, the healing
length is required to be of the same order as the transverse box length.
These quasi-1D condensates can be experimentally realized by loading a
condensate from an  elongated magnetic trap into a 
dipole trap created by a blue detuned Laguerre-Gaussian laser beam 
\cite{kbongs00,dettmer01}.

\subsection{Phase Imprinting}

The method of phase imprinting consists of
passing a short off-resonant laser pulse 
through an appropriately designed absorption plate 
and impinging it on a condensate. In this way one can imprint the
desired phase structure on the condensate and hence create a dark
soliton \cite{dobrek99,dobrek01}.

The Gross-Pitaevskii equation
reduces to an effective one-dimensional nonlinear Schr\"odinger
equation (NLSE)  when the radial frequency is larger than the mean
particle interaction and when the longitudinal dimension of the confining potential 
is much longer than its transverse ones
\begin{eqnarray}
i\hbar \frac{\partial}{\partial t}\Psi(z,t)&=&
\left \{ -\frac{\hbar^2}{2m}\frac{\partial^2}{\partial z^2}+V(z)+
\right. \nonumber \\
&& \left.
\tilde{V}(z,t)+g|\Psi(z,t)|^2 \right \} \Psi(z,t) .
\label{GP1D}
\end{eqnarray}
Here $\tilde{V}(z,t)$ describes the interaction with
the external laser, i.e., denotes the dipole potential generated by the
far detuned laser pulse which acts only in one part of the
condensate, and 
$V(z)$ refers to the time-independent trapping potential 
which remains constant during the whole process.
Let us review here how the phase imprinting can lead to the creation of
a soliton~\cite{dobrek99}.
For a laser pulse duration shorter than the correlation time
of the condensate $\tau_{cor}=\hbar/\mu$, where $\mu$ is the chemical 
potential,  the wave function acquires a local phase factor $e^{-i\phi}$
without changing the condensate's density profile. 
To generate the appropriate phase distribution which leads to a soliton,
it is sufficient to use a potential which acts only on half of the
condensate, e.g.
\be 
\tilde{V}(z,t) = \frac{\hbar \Delta \phi}{ 2} 
\left ( 1+\tanh\left[ \frac{z-z_0}{0.45 l_e} \right] \right)\times f(t)~,
\label{potential}   
\ee
where $f(t)$ is the temporal envelope of the laser pulse normalized
to $\int f(t)dt$=1. This potential imprints a phase
\be
\phi(z) = \frac{\Delta ´\phi}{ 2} 
\left ( 1+\tanh\left[ \frac{z-z_0}{0.45 l_e} \right] \right )~,
\label{phase_i2}
\ee   
where $l_e$ refers to the width of the potential edge, 
which in turn determines the steepness of the imprinted phase gradient at $z_0$. 
Attainable experimental values correspond to
a 10-90\%-absorption width of the phase step. For this reason we
use a factor of 0.45 in Eq.\ref{phase_i2}. 
In accordance to this limit, the experimental values $l_e$ correspond 
to $l_e \ge  2\,\mu$m.
Thus the phase of a dark soliton is composed of two areas of constant phase
connected by a steep gradient.

In an elongated cigar-shaped condensate, 
no matter how accurately the phase-imprinting method is implemented, 
it is not possible to make a standing soliton.  The soliton thus generated will be a moving soliton 
whose speed and depth is directly related to 
$l_e$ and the amplitude of the imprinted phase $\Delta \phi$. 
Figure~\ref{figure1} shows the results of a numerical simulation for
the time evolution of the density and phase
of the condensate within the first millisecond after a phase   
with a  phase gradient of $\Delta\phi =\pi $ and $l_e=2\,\mu$m has
been imprinted.

The imprinted phase profile leads to a velocity field, 
$v_z(z)=(\hbar/m)\partial \phi(z) /\partial z$.
During the evolution on a timescale of the correlation time, 
this velocity field leads to a reduction of density in the region
$\partial v_z /\partial z >0$ ($z>z_0$), whereas in the region 
$\partial v_z /\partial z <0$ ($z<z_0$) the density increases.
After the minimum and the maximum in the density have fully developed, they
begin to back-react significantly, as may be seen 
in the evolution of the phase. The region of the phase gradient 
begins to change and leads to a change in the dynamics of the density 
distribution. The phase gradient splits up into two regions with 
phase gradients  of similar shapes
($\Delta\phi_1\approx \Delta\phi_2\approx \Delta\phi/2$).
The density maximum is connected with one of the phase gradients and
moves approximately with the speed of sound, $c_s=\sqrt{4\pi   
 n_0 a}\hbar/m$, towards negative $z$-values. As time increases the density wave 
broadens
due to dispersion and to the repulsive two-particle interactions.
In contrast, for the minimum propagating towards positive $z$-values, 
the reduced interaction energy results in a compensation 
of the dispersion. This leads to an increase in the steepness of 
the gradient together with a reduction of the width of
the minimum. In this process, a second, 
less pronounced minimum together with
additional density perturbations are created. As may be seen in
Fig.~\ref{figure2}, the created pronounced minimum,  in connection with
the tanh-phase distribution, propagates as a stable solitary wave.

The time scale needed for such a structure to develop is  
approximately given by $\tau_{ds}\approx \tau_{cor}\cdot
(l_e/l_0)$, where $\tau_{cor}$ and $l_0$ correspond to the correlation
time and healing length of the condensate, and the phase step 
connected with the soliton, $\Delta\Phi_2$,
  accounts for approximately   one half of the initially
imprinted phase step $\Delta\phi$. 

The phase imprinting method depends very strongly 
on the width of the potential edge $l_e$ as well as on the value 
of the imprinted phase difference $\Delta\phi$.
For a width much larger than the healing length, $l_e\gg l_0$, 
the time needed for the dark soliton to arise is 
significantly enlarged, and only shallow solitons can be generated. 
For example, for a phase width  of $l_e=5\,\mu$m, but
otherwise identical parameters to those used in Fig.~(\ref{figure1}),
the soliton structure develops only after an evolution time 
of $t_{ev}\approx 15\,$ms.
On the other hand, a phase imprinting with  phases $\Delta\Phi>\pi$ 
but with the same imprinting width $l_e$ leads to a faster development of
soliton structures,  accompanied always by the simultaneous creation 
of multiple solitons.

\subsection{Creation of dark solitons by density engineering}

The possibility of creating solitons in a BEC
by engineering only the phase 
suggests that it should also be possible to create solitons by purely
engineering the density distribution. 
This would  be equivalent to the creation of optical dark solitons
by intensity modulations of a light field propagating 
in a nonlinear medium~\cite{gredeskul89}.

Experimentally it is simple to engineer the density of a condensate.
For instance, one can modify the magnetic trapping potential 
in which the BEC forms with an additional
optical dipole potential of a far blue detuned laser beam,
which is focused to form a thin ``light-sheet'' perpendicular to the
long axis of the cigar--shaped condensate.
The spot size of the focus can easily be chosen to be much 
smaller than the axial size of the condensate. Therefore, in the
Thomas-Fermi limit, the density distribution  of the BEC will be
described by an inverted parabola, except for the region where the laser
focus is applied. This creates a local minimum, $n_{min}$, with
a relative density, $\beta=n_{min}/n_0$, which is controllable by the
laser power. Here $n_0$ is the density at the laser focal position 
for negligible laser power, $P=0$. For a Gaussian laser beam the shape of
the density distribution  in the vicinity of the local minimum will be 
approximately an
inverted Gaussian.  The phase distribution remains constant over the
whole condensate. In order to generate   soliton structures
the dipole potential is non-adiabatically switched off 
while the magnetic trap potential
is kept on.  The phase and density distribution of the condensate adjusts 
to this new   potential by creating  pairs of equal but counter-propagating 
solitons.     Note that the total phase over the condensate is
conserved in this process.  $\beta$ should be $\gtrsim 0.01$ in order to maintain phase
coherence between the two portions of the condensate on either side of the
density notch.

We have numerically simulated the creation of dark solitons in a BEC
in the range of parameters accessible to current experiments
by pure density engineering. For instance, Figure~\ref{figure3}(a)
shows, 10 ms after the non-adiabatic removal of the optical potential,
the density and phase distribution of a $^{23}$Na condensate with
$N=5\times10^6$ atoms.    In this case the magnetic trap has a radial
trapping frequency $\omega_\perp=320\,$Hz and an aspect 
ratio of $\lambda=25$. On the other hand, the optical detuned laser 
has a Gaussian half width at $1/e^2$ of   $W=2\,\mu$m,
and its intensity is assumed to be such that it corresponds to
$\beta=0.9$.  The density profile of the condensate shows three pairs of counter-propagating
dark solitons with different velocities while the phase distribution
depicts the corresponding steep phase gradients in the vicinity of the
solitons.

For a wider laser focus   with respect to the healing length, 
the number of dark soliton pairs increases; 
Fig.~\ref{figure3}(b) shows the situation 10\,ms 
after the switching off of a laser beam with $\beta=0.6$ and $W=12\,\mu$m.
One sees from the figure that after this time 
most of the initial density deformation has already 
been transformed into stable density minima.
For a narrower laser focus with respect to the healing
length it is possible to produce a single pair of solitons \cite{carr00}.

\subsection{Phase and density engineering}

Finally, in this last subsection we briefly 
summarize the results which show 
that  by a proper combination of phase and density engineering 
a single standing dark soliton can be created in a quasi-one dimensional
BEC \cite{carr00}. In this new scenario, a box like confining trap potential
is used for the condensate. 
For a harmonic potential the Thomas Fermi radius scales as 
$N^{1/5}$ and therefore the healing length
scales as $N^{-1/5}$, while for a box-like confinement
the healing length scales as $N^{-1/2}$. Since the size
of the soliton is of the order of twice the healing length it
should be possible, for a box-like confinement with the appropriate
choice of experimental parameters, to dynamically observe the
generated soliton {\it{in situ}}, i.e., without needing to
first expand the condensate. 
The method of density and phase engineering
is simply a combination  of the two above explained methods. First a
density minimum is created by adiabatically ramping the intensity of a
focused  laser into an initially uniform BEC (see Fig~\ref{figure4}). 
The focused laser field
is abruptly switched off and a second far detuned laser pulse of uniform
density is shined on one half of the condensate (phase imprinting). 
   Thus the density minimum acquires the  appropriate phase
distribution.  In this way one can create a single standing dark
soliton.  Variations on this technique allow one to create in a
well-controlled manner asymmetric soliton pairs and various
combinations of larger numbers of solitons.
 
\section{Opening of the trap}

In current experiments using phase imprinting 
the size of the created soliton $\simeq 2\l_0$ is, nevertheless, 
smaller than the diffraction limit of the
wavelength of the imaging radiation,  so the soliton cannot be
observed  {\it in situ} \cite{bur99,dens00}.
To overcome this problem
the trap is suddenly switched off, so that the condensate, 
and therefore the soliton, expands freely for a few milliseconds
($t_{TOF}$) and thus becomes detectable via absorption 
imaging~\cite{bur99}.
In this section we analyze the dynamics associated with the opening of the
trap and the ballistic expansion. We consider a
cigar-shaped geometry with a large aspect ratio. 
The ballistic expansion then occurs principally 
in the transverse direction.
The dynamics related to the opening of the trap is  
complicated, since the abrupt switching off of the trap potential 
modifies not only the density distribution but also the 
phase structure present in the condensate.
Our 3D numerical simulations show that as the condensate expands 
the soliton velocity diminishes  very rapidly while its depth
increases.  Simultaneously a new minimum in the density distribution appears
in the vicinity of the density maximum connected with one of the phase
gradients. This new density minimum observed in the experiments\cite{bur99} travels  
opposite to the soliton direction with a velocity smaller  
than the sound velocity.
Since a soliton can be interpreted as a particle with a negative mass,
by opening the trap the soliton acquires kinetic energy and, therefore,
its velocity decreases until eventually it becomes a 
standing soliton.

A condensate in a quasi-1D trap (infinitely
long cylinder along the axial direction) admits a scaling function for
the wave function\cite{kagan,castin}. Following Ref. \cite{kagan} 
we reexpress the GPE, Eq.(\ref{3DGP}), 
in cylindrical coordinates for the stationary case
\begin{eqnarray}
&&\left\{ - \frac{\hbar^2}{2m}(\Delta_{\rho}+\Delta_z)+ \right. \nonumber \\
&&\left.
\frac{m\omega_{\rho}^2\rho^2}{2}+  
g|\Psi(\rho,z)|^2-\mu \right \}\Psi(\rho,z)=0.
\label{3DGPTI}
\end{eqnarray}
The corresponding stationary soliton solution reads
\be
\Psi_{stat}=\sqrt{\frac{\mu}{g}}\sqrt{1-y^2}\tanh(z\sqrt{1-y^2}))
e^{-i\mu \hbar/t}
\label{statkink}
\ee
where $y=\rho/R_{TF}$ and $z=z/l_0$. We assume here 
for the radial coordinate 
that $\mu\simeq \mu_{TF}$. 
Switching off the trap abruptly 
corresponds to making $\omega_{\rho}$ suddenly zero.
A scaling solution takes the form
\be
\Psi(\rho, z, t)=\frac{1}{b(t)}\Psi_{stat}(\frac{\rho}{b(t)},\frac{z}{b_z(t)}
,\tau)\,e^{i\phi(\rho,z,t)}.
\label{selfsim}
\ee
By substituting Eq. (\ref{selfsim}) into Eq.~(\ref{3DGPTI})
and splitting the real and imaginary part, a solution is found for
an appropriate choice of the phase $\phi$, giving rise to the 
scaling $b_z(t)=\sqrt{1+\omega_{\rho}^2t^2}$. This approach, which is valid for 
$\omega_{\rho}^{-1}\le t_{TOF}\le
\mu/\hbar\omega_{\rho}^2$, predicts a  soliton velocity
\be
v(\tau)=v(0)\frac{\ln(\omega_{\rho}\tau+\sqrt{\omega_{\rho}^2\tau^2 
+1})}{\omega_{\rho}\tau}
\ee
where $\tau=t-t_{open}$ so that $v(0)$ corresponds to
 the velocity of the soliton at the time the
trap is suddenly switched off $(t_{open})$. This scaling 
law agrees very well with the numerical results obtained 
by solving the time dependent GPE, as is shown
in Fig.(\ref{figure6}), as well as with 
the experimental data of Ref.\cite{bur99}.

\section{Interacting solitons}

Solitons propagate in a nonlinear medium without changing their
shape even when they interact with each other.  
The interaction between optical solitons has been 
intensively studied in the propagation of light in monomode optical 
fibers (see Ref. \cite{Kiv98} 
and references therein). Zakharov and 
Shabat demonstrated that homogeneous Bose-Einstein condensates support
multi-soliton solutions and that solitons interact with each other 
like classical particles with a short range repulsive 
interaction \cite{zakharov73}.
The signature of this repulsive interaction 
manifests itself as a negative shift in the position of each 
soliton   after interaction as compared to the position of a single 
free moving soliton. For an untrapped homogeneous 1D condensate   this
shift can be analytically calculated using the inverse scattering method.
In this case the Gross-Pitaevskii equation reduces to 
\be
i\frac{\partial}{\partial t} \Psi(z)=
\left[-\frac{1}{2}\frac{\partial^2}{\partial z^2}+|\Psi(z)|^2\right]\Psi(z).
\label{gp2}
\ee
where $z$ is in units of the healing length and time in units of $\hbar/2g n_0$.

For a repulsive condensate, with the boundary conditions $|\Psi
(z,t)|^2\rightarrow $constant, 
a soliton solution  moving with constant velocity through the condensate
can be reexpressed as \cite{zakharov73,lamb}
\be
\Psi(z,t)=\frac{(\lambda+i\nu)^2+
\exp(2\nu(z-z_0-\lambda t))}{1+\exp(2\nu(z-z_0-\lambda t))} \label{sinsol}
\ee
where $\lambda^2+\nu^2=1$ and $z_0$ is the initial position of the soliton 
at $t=0$. The parameter $\lambda$ characterizes the amplitude and
the velocity of the soliton in units of $c_s$. 
In these units $-1\leq\lambda\leq 1$, where $\lambda=\pm 1$ corresponds to a 
completely filled soliton (zero depth) moving with the speed of
sound, whereas $\lambda=0$ corresponds to a standing soliton.
  
To calculate the spatial shift due to a soliton-soliton interaction
one simply compares each final soliton position to what it would have 
been had it not undergone a collision. 
For two solitons with velocities $\lambda_1$ and $\lambda_2$, 
the resulting shifts are given by \cite{zakharov73,lamb}
\bea
\delta z_1&=& \frac{1}{\nu_1}\log\left[\frac{(\lambda_1-\lambda_2)^2+
(\nu_1+\nu_2)^2}{(\lambda_1-\lambda_2)^2+(\nu_1-\nu_2)^2}\right], \label{s1} \\
\delta z_2&=& \frac{1}{\nu_2}\log\left[\frac{(\lambda_1-\lambda_2)^2+
(\nu_1+\nu_2)^2}{(\lambda_1-\lambda_2)^2+(\nu_1-\nu_2)^2}\right]. \label{s2}
\eea
If the solitons have equal velocities, i.e. $\lambda_1=-\lambda_2$,
the shift is the same for both of them:
\be
\delta z=-\frac{\log[|\lambda_1|]}{\nu_1}.
\ee

Generally speaking, to see how the inhomogeneity due to the trapping potential 
affects the interaction dynamics 
between solitons one has to  turn to numerical 
solutions  of the Gross-Pitaevskii equation. 
To this end, we solve the GPE when two opposite phase gradients  
are imprinted in a cigar-shaped condensate
\be
\phi(z)=\frac{\pi}{2}[\tanh((z-z_1)/l_{e1})-\tanh((z-z_2)/l_{e2})], 
\label{2phases}
\ee
where $z_i$ and $l_{ei}$ denote the positions of the phase gradients and
the width of the potential edge respectively.
Thus, at $t=0$  the wavefunction of the condensate reads
\be
\Psi(z,t=0)=e^{-i\phi(z)} \psi(z),
\ee 
where $\psi(z)$ is the ground state solution of the Gross-Pitaevskii 
equation. After a time of the order of the correlation time, 
such a phase distribution generates two counter propagating solitons 
(two notch planes located at
$z_1$ and $z_2$) moving with velocities ($\lambda_1=-\lambda_2$), 
together with two counter propagating density waves 
(density maxima) that move with the speed of sound.
The soliton positions are monitored by following the density notch 
in the condensate. Figs. \ref{figure7} and \ref{figure8} display 
the results of a full 3D calculation corresponding to the experimental
conditions discussed in \cite{bur99}, where the soliton planes are
initially separated by 20 and 35 microns respectively. 
Due to the change in the background density of the condensate, the
solitons first accelerate, and then cross each other.  
In Fig. \ref{figure7} for the case 
of an initial separation of
20 microns the two solitons overlap strongly 
during the collision, which lasts for approximately 1 ms. 
In Fig. \ref{figure8} for an initial separation of 35 microns, the
solitons collide with such a high velocity, that the short interaction time 
makes the position shifts negligible.

In spite of the 3D character of the elongated trap
one can still use the analytical results 
\cite{zakharov73} from the homogeneous 1D case by applying a 
local density approximation where the condensate density is considered 
constant in the region of the soliton plane. This approximation 
is valid away from the the edges of the trap.
Inserting the parameters used in the 
Hannover soliton experiment \cite{bur99} for $^{87}$Rb  with N=$10^5$, a=5.7 nm, 
and an initial $v_{soliton}=3.3 \mu m/ms$, and assuming an effective
cross section area of $S=25 \mu m^2$,  the predicted shift in the position
of the solitons is $\delta z=-0.1 \mu m$. This calculated value 
is in agreement with the numerical simulations. 
Such a small positional shift cannot be detected experimentally. In order to 
obtain a shift of the order of $10 \mu m$ the velocity difference should
be $|v_1-v_2| \sim 10^{-6} c_s$.

Finally, in Fig.\ref{figure9} we display the results (1D) for
the collision between two dark solitons created with 
phase gradient of {\it $\l_e=10$}. The initial
velocities are very small $(v\sim 0.05 c_s)$. Thus the interaction time becomes
very large.  The solitons are clearly seen 
to bounce off each other like classical particles undergoing an
elastic collision.
The shift in the positions is, however, still very small. 
Here we have chosen the density to be   $4\times 10^{13} {\rm cm}^{-3}$
which again corresponds to a condensate with the effective
cross sectional area $S=25 \mu {\rm m}^2$. 

In order to unambiguously detect experimentally 
a signature of the interaction between solitons in condensates,
one should create two solitons with very similar velocities 
propagating in the same direction.  
The solitons will thus interact for a long time. 
From Eqs. (\ref{s1}) and (\ref{s2}) it is clear 
that this scenario can amount to an arbitrarily large shift. This
requires, however, a very long condensate. Such condensates are
becoming experimentally available~\cite{dettmer01,ketterle} in reduced
geometries. On the other hand, 
this drawback could be removed by creating the condensate in a ring geometry. 
Another possibility is to create solitons in a hard wall geometry, 
as for example   in a hollow blue-detuned laser beam with laser light
sheet endcaps, so that the solitons will reflect from the ends without changing 
their form, since the wavefunction at the endcaps acts as a 
pinned soliton~\cite{reinhardt1}.

\section {Stability of dark solitons for negative scattering lengths}

Recent experiments concerning the use of Feshbach resonances
\cite{scat00} to change both the magnitude and the sign of
the scattering length of condensates with alkali atoms 
offers a new range of phenomena to study. In particular,
a carefully controlled study of the outstanding problem of
collapse of the condensate for negative scattering length becomes possible~\cite{roberts01}.
As already mentioned in the previous
sections,   the form of a soliton depends on the sign of the scattering
length. For positive scattering lengths the stable soliton solution
is a density notch, i.e, a density minimum; for sufficiently strong
negative scattering 
lengths the stable solution is a bright soliton which is a density
peak. 

  The possibility of changing the scattering length from positive
to negative values opens the question of the stability of dark 
solitons in attractive condensates.
Let us first reconsider the stability of dark solitons 
when adiabatically changing the scattering length.
Since a soliton is a
particular solution of the Gross-Pitaevskii equation for a
well defined scattering length, it is reasonable to assume an adiabatic change 
of the scattering length will change the
velocity    and depth of the soliton gradually. 

On the contrary, an abrupt change in
the condensate's scattering length will destroy the soliton.  
For negative scattering lengths the
physical and  mathematical situation changes dramatically. If we only
consider low densities and neglect three-body recombination -- which
becomes important at high densities and produces additional
kinetic energy -- in the absence of any solitons, the instability of
the condensate is seen  as a collapse of the condensate's
wavefunction.  
This is shown in Fig. \ref{figure10}, where one can see a
shrinking cloud as time proceeds and where no solitons
are present. In these simulations 
we have used a small $^{87}$Rb condensate with 5000 atoms in the 
same cigar-shaped trap previously discussed. We 
change the scattering length from its initial value $a=5.7 $nm to $a\rightarrow -0.1$a.
In the presence of a dark soliton, (Fig.~\ref{figure11}) the scenario changes dramatically. 
The soliton splits the cloud into two separate parts which independently continue to
collapse. A direct consequence of non-adiabatically 
changing the sign of the scattering length
in the presence of a soliton   is the creation of a large number 
of density waves.
This effect speeds up the collapse of the wave function
because of the local increase in the density. A more
careful study which takes into account three-body recombination
 processes is needed to investigate the dynamics for longer times.

\section{conclusions}
We have discussed the generation, evolution, and interaction of dark solitons
in matter waves. We have first reviewed different approaches to generate
standing or moving dark solitons in one component condensates. The interaction
dynamics between dark solitons have also been addressed and we have
discussed under which circumstances the interaction can be observed
experimentally. 
We conclude that in   present experiments using cigar-shaped condensates with a large
aspect ratio, a conclusive signature of the soliton interaction 
cannot be observed. However, by using other geometries, such 
  as quasi-1D or toroidal condensates, the interaction could be
unambiguously detected experimentally.     A stationary wave in
the form of a density notch or peak, even if it
moves with less than the speed of sound, cannot truly be called a
soliton until it is demonstrated that it interacts as one, since that
is the defining characteristic which gives solitons their
particle-like nature.

Finally, we have discussed the 
stability of dark solitons to sudden changes of the sign and value of
the scattering length. We find that the presence of a dark soliton
can be unambiguously detected by the radical change in the dynamics of
a collapsing cloud when a Feshbach resonance is used to tune the scattering length negative: the
cloud splits in two.

\acknowledgements
  
We acknowledge support from  SFB407 of the 
{\it Deutsche Forschungsgemeinschaft}, the  National Science Foundation 
Grant CHE97-32919, PESC BEC2000+ and EPSRC. Discussions with K. Bongs, M. Lewenstein, 
L. Santos, J. Brand, W. Ertmer and in particular G. Shlyapnikov (sec
III) are gratefully acknowledged.

\begin{figure}
\caption{
Numerical simulations (1D) showing the 
time evolution of the density profile of a 
$^{87}$Rb condensate with $N=10^5$. 
The static trap has a frequency $\omega_z=2\pi\times 14$  Hz.
The density profile $n(z)$ and the phase
are depicted respectively in (a) and (b) within the first 
millisecond after a phase-imprint with $\Delta \phi =\pi$
and $l_e=2\,\mu$m has been implemented.}   
\label{figure1}
\end{figure}

\begin{figure}
\caption{
Time evolution of the density profile $n(z)$ 
and the phase $\phi(z)$ for the first 5 milliseconds after
a $\pi$-phase has been imprinted (otherwise the same parameters
as in Fig~\ref{figure1}).}   
\label{figure2}
\end{figure}

\begin{figure}
\caption{     
Evolution of pairs of counter propagating dark solitons 
created by a non-adiabatic change of the BEC trapping potential.
The parameters here correspond to a Na-condensate with $N=5\times10^6$ atoms
in a magnetic trap of the clover-leaf type with a radial trapping
frequency of  $\omega_\perp=320\,$Hz and an aspect ratio of $\lambda=25$:
(a) $W=2\,\mu$m, $\beta=0.9$  and (b) $\beta=0.6$ and $W=12\,\mu$m.
 In the same figure, density profile and phase gradients are
depicted.    Note that the phase in (b) has been plotted modulo
$2\pi$.} 
\label{figure3}
\end{figure}

\begin{figure}
\caption{(a) A combination of a box-like potential and a tightly focused,
blue-detuned laser beam is used to engineer the density. (b) The resulting
wavefunction is phase engineered with a second, far-detuned laser beam,
resulting in an initial state that resembles very much a standing dark soliton.}
\label{figure4}
\end{figure}

\begin{figure}
\caption{Velocity of the soliton outside the trap versus observation
time. The soliton is imprinted at $t=0$ and the trap is suddenly
removed after $t=4$ ms. The curve shows
the analytical scaling law, and the dots correspond to the values of the
soliton velocity obtained from numerical 3D simulations.}
\label{figure6}
\end{figure}

\begin{figure}
\caption{Interaction between solitons in a cigar shaped condensate.
The solitons are initially created by the method of phase imprinting
with an initial separation of 20 microns. Here N=$10^5$ and a=5.7 nm.}
\label{figure7}
\end{figure}

\begin{figure}
\caption{The same parameters as in Fig. 6, with the solitons 
initially separated by 35 microns. At the time of the collision their
velocities are now higher, hence the interaction time shorter. 
}   
\label{figure8}
\end{figure}

\begin{figure}
\caption{The interaction between solitons in the 1D case. The density waves
created by the phase-imprinting are moving with the speed of sound and
are reflected at the condensate boundary (not shown
in the figure) which eventually results in crossing waves at $t=12 ms$.} 
\label{figure9}
\end{figure}

\begin{figure}
\caption{The scattering length is changed at $t=2ms$ from
  $a$ to $-0.1 a$. The cloud, represented here as the integrated
  density as a function of $z$, is starting to collapse at the onset 
of the negative scattering length.}   
\label{figure10}
\end{figure}

\begin{figure}
\caption{Same situation as in Fig. \ref{figure10} with the presence of a
dark soliton. The soliton splits the cloud into two parts which
independently start to collapse.}   
\label{figure11}
\end{figure}


\end{document}